\documentclass[conference]{IEEEtran}
\IEEEoverridecommandlockouts
\usepackage{cite}
\usepackage{amsmath, amssymb,amsfonts}
\usepackage{algorithmic}
\usepackage{graphicx}
\usepackage{textcomp}
\usepackage{xcolor}
\def\BibTeX{{\rm B\kern-.05em{\sc i\kern-.025em b}\kern-.08em
    T\kern-.1667em\lower.7ex\hbox{E}\kern-.125emX}}
\begin{document}

\title{Patient Domain Supervised Contrastive Learning for Lung Sound Classification Using Mobile Phone\\
\thanks{This work was supported by the National Research Foundation of Korea (NRF) funded by the Korean Government under Grants RS-2023-002084912 and 2020M3C1B8081320. (Corresponding author: Seong-Eun Kim)}}

\author{\IEEEauthorblockN{Seung Gyu Jeong and Seong-Eun Kim}
\IEEEauthorblockA{\textit{Dept. Applied Artificial Intelligence} \\
\textit{Seoul National University of Science and Technology}\\
Seoul, South Korea \\
wa975@naver.com, sekim@seoultech.ac.kr}
}

\maketitle

\begin{abstract}
Auscultation is crucial for diagnosing lung diseases. The COVID-19 pandemic has revealed the limitations of traditional, in-person lung sound assessments. To overcome these issues, advancements in digital stethoscopes and artificial intelligence (AI) have led to the development of new diagnostic methods. In this context, our study aims to use smartphone microphones to record and analyze lung sounds. We faced two major challenges: the difference in audio style between electronic stethoscopes and smartphone microphones, and the variability among patients. To address these challenges, we developed a method called Patient Domain Supervised Contrastive Learning (PD-SCL). By integrating this method with the Audio Spectrogram Transformer (AST) model, we significantly improved its performance by 2.4\% compared to the original AST model. This progress demonstrates that smartphones can effectively diagnose lung sounds, addressing inconsistencies in patient data and showing potential for broad use beyond traditional clinical settings. Our research contributes to making lung disease detection more accessible in the post-COVID-19 world.
\end{abstract}


\section{Introduction}
Lung sound classification has gained prominence in healthcare, particularly in diagnosing and monitoring respiratory disease. 
Accurate lung sound classification leads to early disease detection, timely intervention, and improved patient outcomes. 
In the post-COVID-19 era, its importance has been further enhanced as respiratory health has become a paramount concern \cite{one}. In addition, the method of a doctor auscultating lung sounds face-to-face in the COVID-19 situation had several difficulties because doctors also had a risk of infection. 
These difficulties have further increased the need for non-face-to-face diagnosis. 

For non-face-to-face diagnosis, the development of diagnostic devices and diagnostic algorithms is inevitable. With the recent development of electronic stethoscopes, lung sounds can be easily collected\cite{two}, which has greatly helped develop artificial intelligence technology that is essential for non-face-to-face diagnosis. Research on the development of various artificial intelligence technologies using collected lung sound is actively being conducted\cite{three,four,five}, and among them, the results using the Audio Spectrogram Transformer (AST) model stand out.
 
We do not stop there, and we intend to expand the possibility of such non-face-to-face lung sound diagnosis to mobile phones. Through direct cooperation with hospitals, we collected the recorded lung sound of the mobile phone's internal microphone and proved that the artificial intelligence model can be classified sufficiently. In addition, since the amount of data collected on mobile phones is not large, we studied how to increase the performance by combining it with the vast amount of data collected with the existing electronic stethoscope.
 
We define the problem we face from two perspectives and suggest ways to solve it. First, we tried to reduce the difference in data characteristics between two measuring devices, an electronic stethoscope, and a mobile phone. Second, we tried to reduce the difference in data characteristics that each patient has. To solve these two problems, patient domain supervised contrastive learning was presented in this paper, and experiments were conducted in combination with the AST model. The results were compared through experiments when the Patient Domain Supervised Contrastive Learning (PD-SCL) method was applied, when it was not, and when a different method of domain adaptation was applied.

\section{Related Work}

\subsection{Lung Sound Classification}

Recent advancements in lung sound classification have leveraged deep learning to address challenges in detecting abnormal lung sounds, especially in scenarios with limited data. The development of RespireNet showcased a significant leap forward, demonstrating a deep neural network's capability to identify abnormal lung sounds with high accuracy in constrained data settings \cite{six}. Similarly, the work on pretraining respiratory sound representations using metadata and contrastive learning illustrated how leveraging additional data attributes can enhance model performance \cite{seven}. The introduction of Patch-Mix Contrastive Learning with the Audio Spectrogram Transformer marked another innovative approach, further refining respiratory sound classification by incorporating sophisticated data augmentation and representation learning techniques \cite{eight}. Furthermore, the concept of stethoscope-guided supervised contrastive learning has been explored to facilitate cross-domain adaptation, underscoring the potential of contrastive learning mechanisms to improve classification accuracy across varied recording environments \cite{nine}.

\subsection{Domain Adaptation}

Domain adaptation, a crucial aspect of machine learning, aims to enhance model performance across varying domains. Early strategies, focusing on feature and instance-based methods, evolved into sophisticated deep learning approaches, especially in unsupervised and semi-supervised scenarios, to address significant challenges like cross-domain discrepancy and scarcity of labeled data in target domains \cite{ten,twelve}. This field has seen extensive applications in areas like computer vision and natural language processing, underscoring its importance in adapting models to diverse and evolving data landscapes \cite{thirteen}. As the domain continues to mature, ongoing research is increasingly focusing on robust model development and integrating unsupervised learning techniques to further bridge domain gaps \cite{fourteen}.

\subsection{Supervised Contrastive Learning }

Supervised Contrastive Learning, an advanced twist on traditional contrastive learning, leverages labeled data to enhance learning efficacy by focusing on distinguishing between similar (positive) and dissimilar (negative) examples. This approach, evolving from its unsupervised counterpart, utilizes carefully chosen loss functions and pair selection methods to strengthen feature representations, particularly in complex tasks like computer vision and natural language processing \cite{fifteen,sixteen}. While offering significant improvements over conventional methods in handling complex data structures, it faces challenges like efficient negative sampling and balancing, paving the way for future research to optimize these aspects \cite{seventeen}.

\section{Method}

\subsection{AST}

In the field of respiratory sound analysis, the obstacle of scarce training data has been mitigated by the use of models pre-trained on large databases such as ImageNet (IN) and AudioSet (AS) \cite{eighteen}. 
The AST model's ability to process long-range dependencies makes it particularly adept at handling complex audio scenes where events or sounds of interest are dispersed over time. Its scalable architecture, which can be adjusted based on the dataset and task complexity, further contributes to its versatility and effectiveness in a wide range of audio analysis applications. 

\subsection{Domain Adversarial Training}

In order to deal with the variation in data distribution across different recording sources, we integrated a Domain Adversarial Training (DAT) methodology inspired by \cite{nineteen}. This approach includes a dual loss system: the cross entropy loss (LCE) and the domain adversarial loss (LDA), articulated as follows:
\begin{align}
    L_{CE} &= -\sum_{i=1}^{n} y_i \log(\hat{y}_i), & L_{DA} &= -\sum_{i=1}^{n} d_i \log(\hat{d}_i),
\end{align}
where $y_i$ and $d_i$ are the class and domain labels, respectively, and $\hat{y}_i$ and $\hat{d}_i$ are the predicted probabilities. The total loss is given by
\begin{equation}
    L_{total} = L_{CE} + \lambda L_{DA},
\end{equation}
where $\lambda$ is the domain regularization parameter, set to 0.2 in our experiments.

\subsection{PD-SCL}
The Patient Domain Supervised Contrastive Learning (PD-SCL) is a novel approach developed to improve lung sound classification by accounting for variability across data sources and patient-specific variations. 

The PD-SCL method involves several key steps in its implementation:

\begin{enumerate}
    \item \textbf{Feature Normalisation:} Features are normalized using an L2 norm to ensure consistency of the scale.
    \item \textbf{Similarity Matrix Computation:} A similarity matrix is computed by performing a dot product of the normalized feature vectors, scaled by a temperature parameter. This matrix quantifies the similarity between each pair of samples in feature space.
    \item \textbf{Pairwise Mask Generation:} Masks for positive and negative pairs are generated based on labels, patient IDs, and domain IDs. Positive pairs have the same label but come from different patients or domains, while negative pairs have different labels.
    \item \textbf{Log-Sum-Exp Trick:} Applied to the similarity matrix for numerical stability, especially when dealing with large exponents.
    \item \textbf{Summation of Similarities:} The sum of negative similarities is calculated and normalized by the number of pairs to ensure balanced influence.
    \item \textbf{Loss Calculation:} The loss contrasts the sum of positive similarities with the sum of negative similarities, promoting effective discrimination between positive and negative pairs.
    \item \textbf{Total Loss:} The average of the individual losses across the batch represents the overall PD-SCL.
\end{enumerate}

The similarity between two data samples $i$ and $j$ is calculated using the formula:

\begin{equation}
    \text{Sim}(i, j) = \frac{f_i^T f_j}{\tau},
\end{equation}
where $f_i$ and $f_j$ are the feature vectors extracted from a neural network model for these data points, and $f_i^T$ represents the transpose of $f_i$. The temperature parameter $\tau$ scales this dot product, adjusting the similarity values' intensity.

The PD-SCL loss ($L_{PD-SCL}$) ( for a mini-batch of $N$ samples is defined as:
\begin{equation}
    \text{$L_{PD-SCL}$} = \frac{1}{N} \sum_{i=1}^{N} -\log\left(\frac{\sum_{j \in P(i)} \exp(\text{Sim}(i, j))}{\sum_{k=1, k \neq j}^{N} \exp(\text{Sim}(i, k))}\right),
\end{equation}
where positive samples (\(j \in P(i)\)) for sample \(i\), are defined as those that belong to the same class as $i$ but originate from different patients or domains. Conversely, negative samples consist of all other samples except positive samples(\(k \neq j\)). 

The total loss function is a combination of cross-entropy loss $L_{CE}$ and PD-SCL loss $L_{PD-SCL}$ :
\begin{equation}
    L_{total} = L_{CE} + \lambda L_{PD-SCL},
\end{equation}
where $\lambda$ is a balancing parameter. In this study, we set $\lambda = 0.5$ and $\tau = 0.5$. The PD-SCL method significantly reduces both domain- and patient-specific variability in lung sound data. By effectively fitting data distributions, it improves the model's ability to generalise across different datasets and patient profiles, a critical factor for reliable lung sound classification in clinical and real-world settings.

\section{Result}

\begin{table}[t]
\caption{Data Information}
\label{tab:data}
\begin{center}
\begin{tabular}{|c|c|c|}
\hline
\textbf{Condition}&\multicolumn{2}{|c|}{\textbf{Device}} \\
\cline{2-3} 
 & \textbf{\textit{Stethoscope}}& \textbf{\textit{iPhone}} \\
\hline
Normal& 4376& 1410\\
\hline
Abnormal (Crackle, Wheeze, Both)& 4348& 1291\\
\hline
Total& 8724&2701\\
\hline
\end{tabular}
\label{tab:data_updated}
\end{center}
\end{table}

\subsection{Dataset Description }

Our study leveraged a unique dataset of lung sound recordings obtained through two different devices. One set of recordings was made using a digital stethoscope (specifically, a 3M Littmann digital stethoscope) and the other set was recorded from mobile phones (utilizing the iPhone's recording function). These recordings were conducted by medical professionals on pediatric patients at Woorisoa Children's Hospital in Seoul, Korea, ensuring a high standard of data quality and relevance. The dataset includes a total of 8,724 recordings from the digital stethoscope, comprising 4,376 normal and 4,348 abnormal lung sounds. Meanwhile, the mobile phone dataset consists of 2,701 recordings, with 1,410 classified as normal and 1,291 as abnormal. Together, the combined dataset features 11,425 recordings, encompassing 5,786 normal and 5,639 abnormal sounds.
 The stethoscope dataset was collected from 342 patients, while the mobile phone dataset was collected from 63 patients. It is noteworthy that all patients recorded with mobile phones also have corresponding stethoscope recordings, with additional patient data exclusive to the stethoscope dataset. For cross-validation purposes, we divided the data into 5 folds, using both stethoscope and mobile phone recordings for training, but using only mobile phone recordings for validation as detailed in Table~\ref{tab:data}. Leave-subject-out cross-validation was applied to the patients used in the test data so that they were not used in the training process

\subsection{Data Preprocessing}

To prepare the data for input into the Audio Spectrogram Transformer (AST)\cite{eighteen} model, all recordings were standardized to 8-second intervals at a sampling rate of 16kHz. Recordings shorter than 8 seconds were extended using repeat padding, while longer recordings were truncated. The audio waveform was then converted into a sequence of 128-dimensional log Mel filterbank (Fbank) features using a window size of 25 ms and an overlap size of 10 ms. Standard normalization was applied to the spectrograms with a mean of -4.27 and a standard deviation of 4.57.

\subsection{Evaluation Metric}

To evaluate the classification performance, we adopted the evaluation metric from the ICBHI 2017 open dataset. Our primary metric was the score, which was calculated as the average of specificity (Sp) and sensitivity (Se). Specificity and sensitivity are defined as:

  \[ Sp = \frac{C_n}{N_n}, \] 
  
  \[ Se = \frac{C_c + C_w + C_b}{N_c + N_w + Nb}, \]
  
  \[ Sc = \frac{Sp + Se}{2}, \]
where $\{C_i\}$ and $\{N_i\}$ are the number of correctly classified and total samples in class $i$, and $i \in \{\text{normal}, \text{abnormal}\}$. For the cases of crackle, wheeze, and both, we considered them all as abnormal and used them as a two-class classification.

\subsection{Performance Comparison}
We evaluated the performance of each method over 5 folds, focusing on specificity (Sp), sensitivity (Se), and combined score (Sc). The following table summarises the performance metrics for each method:

\begin{table}[t]
\caption{Performance between Different Methods}
\label{tab:freq}
\begin{center}
\begin{tabular}{|l|c|c|c|}
\hline
\textbf{Method} & \textbf{Sp (\%)} & \textbf{Se (\%)} & \textbf{Sc (\%)} \\
\hline
AST - Finetuning (Mobile data only) & 85.7 & 82.5 & 84.1 \\
\hline
AST - Finetuning (Combined Data) & 86.3 & 81.5 & 83.9 \\
\hline
DAT & \textbf{92.5} & 74.8 & 83.7 \\
\hline
\textbf{Our (PD-SCL)} & 89.9 & \textbf{82.7} & \textbf{86.3} \\
\hline
\end{tabular}
\label{tab:freq_updated}
\end{center}
\end{table}

\begin{figure}
    \centering
    \includegraphics[width=0.4\textwidth]{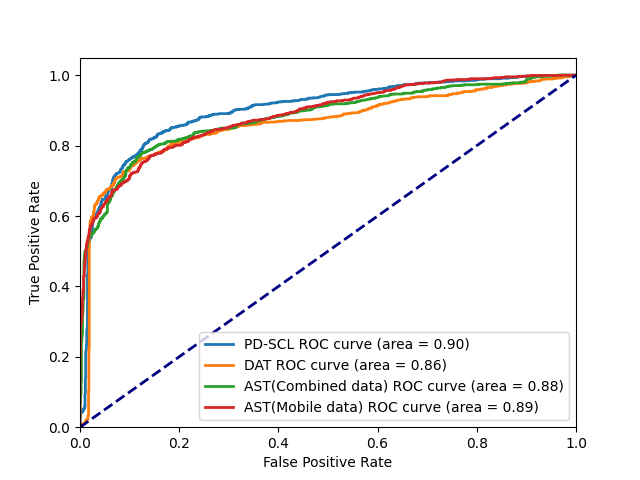}
    \caption{ROC Curve and AUC for each model}
    \label{fig1}
\end{figure}

In our study, we systematically evaluated the efficacy of the Audio Spectrogram Transformer (AST) model under various conditions. Our objective was to assess how different dataset compositions and learning methodologies impact the model's performance in processing audio data.


\noindent\textbf{AST (Mobile Data Only):}
Fine-tuned AST model using only mobile phone data. 

\noindent\textbf{AST (Combined Data):}
Fine-tuned AST model using both mobile phone, electronic stethoscope data. 

\noindent\textbf{DAT:}
Domain Adaptation Training (DAT) was applied, using both mobile phone, electronic stethoscope data.

\noindent\textbf{PD-SCL:}
Our novel learning method, Patient Domain Supervised Contrastive Learning (PD-SCL), was utilized, incorporating a comprehensive dataset similar to the previous cases.

The results of Table~\ref{tab:freq} show that using mobile data alone outperforms both without domain adaptation or using DAT methods. However, the PD-SCL method performs best when both data are combined.
Our method proves the hypothesis that the performance of lung sound data classification can be significantly improved by mitigating the imbalance in the data caused by instrument variability and patient-specific factors. Our results highlight the important role of domain adaptation techniques in improving the accuracy and reliability of lung sound analysis by minimising data variation between measurement sources and patient characteristics.
We have also visualised the ROC curves and AUC for each model in Figure~\ref{fig1}, which demonstrates the ability of our models to classify abnormal sounds more efficiently, rather than being biased towards normal sounds.
 
\section{Conclusion}

The PD-SCL method, a key innovation in our study, has demonstrated exceptional robustness and generalisability across different patient and domain characteristics. This approach significantly outperformed other domain adaptation techniques in validation results, especially in the challenging context of lung sound classification. A key finding of our research is demonstrating how patient-specific data variations can affect the performance of medical data analysis. By implementing PD-SCL, we have successfully developed a robust method for both patient-specific and domain-specific distributional differences. This advance is particularly important in the medical field, where individual patient differences can have a significant impact on diagnostic accuracy. The proposed PD-SCL method marks a significant step forward in domain adaptation for medical data. It addresses the gap between different types of data sources, such as high-quality stethoscope recordings and variable mobile phone audio, and effectively bridges these variations. Our results demonstrate that this method can improve performance even when faced with the inherent variability and smaller dataset sizes of mobile phone recordings.

\bibliographystyle{IEEEtran}
\bibliography{ref}

\end{document}